\documentclass[
    ,final            
  ]
  {aipproc}

\layoutstyle{8x11single}

\begin{document}

\title{Transition nucleon resonance electrocouplings from CLAS data on
$\pi^{+}\pi^{-}p$ electroproduction off protons}


\author{Victor I. Mokeev}{
  address={Thomas Jefferson National Accelerator Facility, Newport News, 
  Virginia 23606, USA},
  email={mokeev@jlab.org},
   altaddress={Skobeltsyn Nuclear Physics Institute at Moscow State University,
    Moscow, Leninskie gory 119889, Russia}
  }


\begin{abstract}
{Electrocouplings of excited proton states with masses less than 1.8 GeV 
were determined for the first time from the CLAS data on $\pi^+\pi^-p$ 
electroproduction at photon virtualities $Q^2$ $<$ 1.5 GeV$^2$. Electrocouplings were obtained 
from a combined fit of all available observables
within the framework of a phenomenological reaction model.}
\end{abstract}

\maketitle


\section{Introduction}
Studies of nucleon resonance electrocouplings $\gamma_{virt}NN^*$ from 
$ep \rightarrow e' \pi^+\pi^-p $ channel  represent an important part of
the $N^*$ Program with the CLAS detector \cite{Bu09}. The $N\pi$ and  
$\pi^+\pi^-p$ exclusive electroproduction channels are major contributors in
the resonance excitation region with entirely different non-resonant mechanisms. Therefore, they
provide independent information on  $\gamma_{virt}NN^*$
electrocouplings, that is vital for the reliable extraction of these fundamental
quantities. Comprehensive information on $\gamma_{virt}NN^*$ electrocouplings 
of various excited proton states in a wide range
of photon virtualities $Q^2$ gives access to active degrees of freedom in the 
$N^*$ structure at various
distances. This allow us to explore non-perturbative strong interaction mechanisms, that are
responsible for baryon formation. Preliminary results on the 
$Q^2$-evolution
of  $\gamma_{virt}NN^*$ electrocouplings for excited proton states 
with masses less than 1.8 GeV are reported. 
They were obtained for the first time
from  $\pi^+\pi^-p$ electroproduction off protons \cite{Ri03,Fe09}.

\section{Analysis approach}
\label{an}
The CLAS data on
$\pi^+\pi^-p$ electroproduction \cite{Ri03,Fe09} for the first time provided information 
on nine independent one-fold differential and
fully integrated cross sections in each bin of $W$ and $Q^2$ in a mass range 1.31 $<$ $W$ $<$ 2.1 GeV 
and at photon virtualities from 0.25 to 1.5 GeV$^2$. 
Analysis of these data allowed us to establish 
all essential mechanisms 
contributing
to $\pi^+\pi^-p$  electroproduction in this kinematical area. 
The presence and strengths of the contributing
$\pi^+\pi^-p$ electroproduction mechanisms was established
by studying the kinematical dependencies in differential cross sections 
and their correlations in a variety of available observables. 
This resulted in the development of  meson-baryon reaction model JM  
for the description 
of $\pi^+\pi^-p$ electroproduction off protons \cite{Mo09,Mo06}. 
The primary objective is to determine 
$\gamma_{virt}NN^*$ electrocouplings and $\pi\Delta$ and $\rho p$ 
partial hadronic decay widths of $N^*$'s from a combined fit of all
available observables. The model incorporates the full 
$\pi^+\pi^-p$ production amplitude of the $\pi^-
\Delta^{++}$, $\pi^{+} \Delta^{0}$, $\rho p$, $\pi^{+}D^{0}_{13}(1520)$, $\pi^{-}P^{++}_{33}(1600)$,
$\pi^{+}F^{0}_{15}(1685)$ isobar channels and direct double pion
production mechanisms, that account for all resonant and non-resonant 
partial waves combined.  Direct double pion
production mechanisms describe processes when the final $\pi^+\pi^-p$ state is created without formation of
unstable hadrons in intermediate states. 
Direct $2\pi$ production mechanisms, required by general unitarity 
condition \cite{Ait72}, were established for the first time in the 
analysis of CLAS   
$\pi^+\pi^-p$  electroproduction data. Implementation of direct $2\pi$ production 
processes into the
JM model allowed us to move beyond the isobar approximation 
for the description of $\pi^+\pi^-p$ 
electroproduction off protons. The JM model incorporates all $N^*$'s with
masses less than 2.0 GeV that couple to the two pion channel. 
Resonant amplitudes contribute to $\pi \Delta$
and $\rho p$  isobar channels, while the other  
isobar channels contain only non-resonant 
mechanisms. Breit-Wigner
parametrization for resonance amplitudes, employed in the JM model (the regular 
Breit-Wigner ansatz), 
is described in \cite{Ri00,Mo01}. However, this ansatz doest not satisfy 
unitarity
condition in a case, when various $N^*$ states could be mixed 
in a dressed
resonance propagator.
Therefore, in the 2010 JM model version the resonant amplitudes were parametrized 
using a unitarized Breit-Wigner ansatz, that was initially proposed in \cite{Ait72}.
 This ansatz was modified to make it consistent with resonant amplitude 
 parametrization adopted in the JM model. The unitarized Breit-Wigner ansatz
allowed us to account for transitions between various $N^*$ states in the 
dressed resonance s-channel propagator  and to impose restrictions on resonant amplitudes required
by general unitarity condition. Non-resonant amplitudes incorporated into the
 JM model were presented
in the papers \cite{Mo09,Mo06}. 

A reasonable description of the $\pi^+\pi^-p$ electroproduction channel was 
achieved, allowing us to separate resonant and
non-resonant contributions to the measured cross sections, which is needed 
for the evaluation of $\gamma_{virt}NN^*$
electrocouplings.

\section{Fitting procedure}
The $N^*$ electrocouplings and their $\pi \Delta$ and $\rho p$ partial 
hadronic decay widths  were obtained from fits to 
the $\pi^+\pi^-p$ electroproduction data \cite{Ri03,Fe09}  
within the framework of JM model \cite{Mo09,Mo06}. 
These resonance parameters, as well as parameters describing 
the non-resonant mechanisms in the JM model were varied simultaneously in 
the $\chi^2$ minimization.
In this way we accounted for correlations of resonant and non-resonant 
contributions. 
For each trial set of computed cross sections 
 the $\chi^2$/d.p. value
 was estimated in point by point comparison between measured and computed 
 nine one-fold
 differential cross sections in all bins of $W$ and $Q^2$ covered by
 measurements. 
  
\begin{figure}[htp]
\includegraphics[width=7cm]{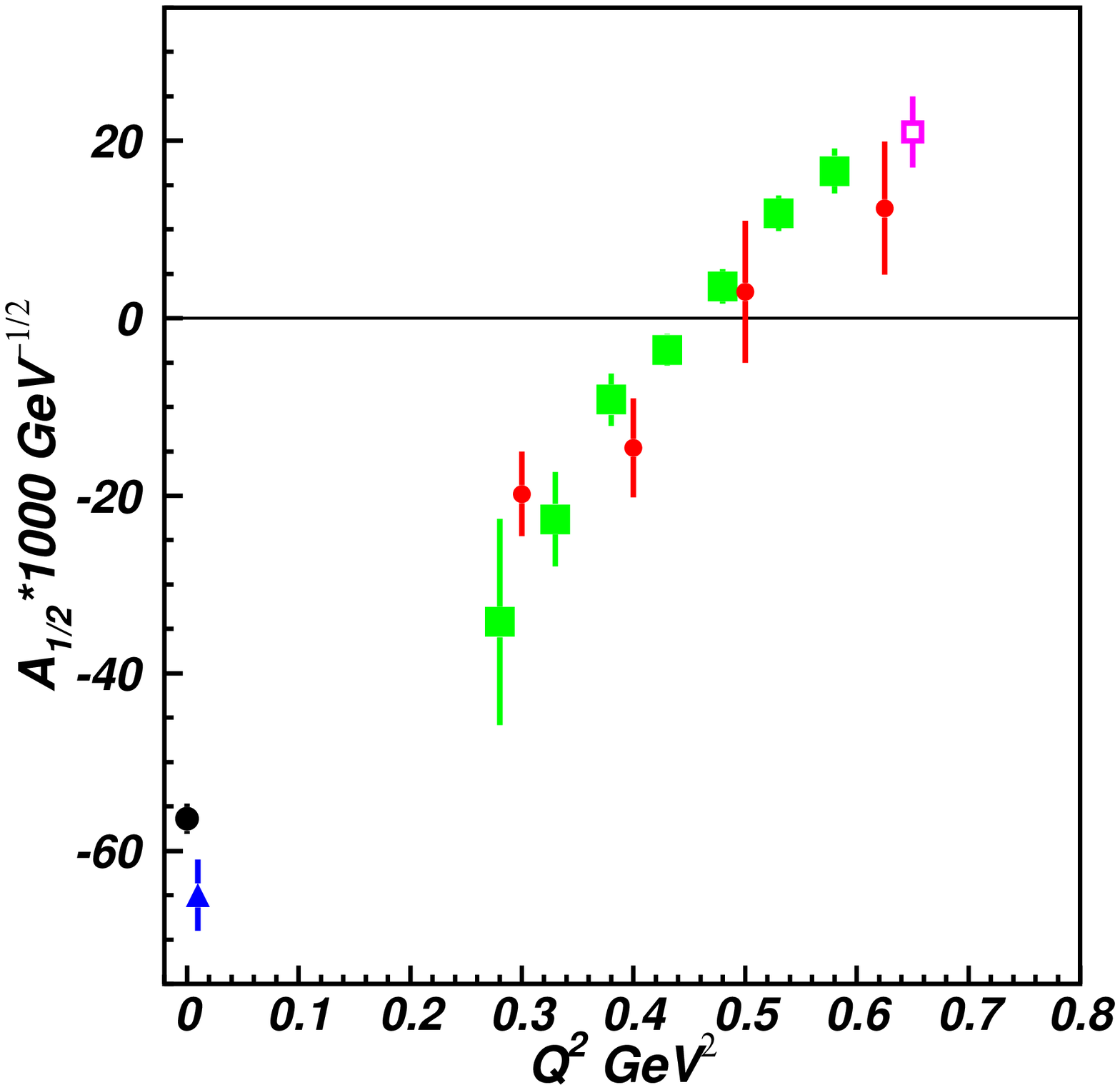}
\includegraphics[width=7cm]{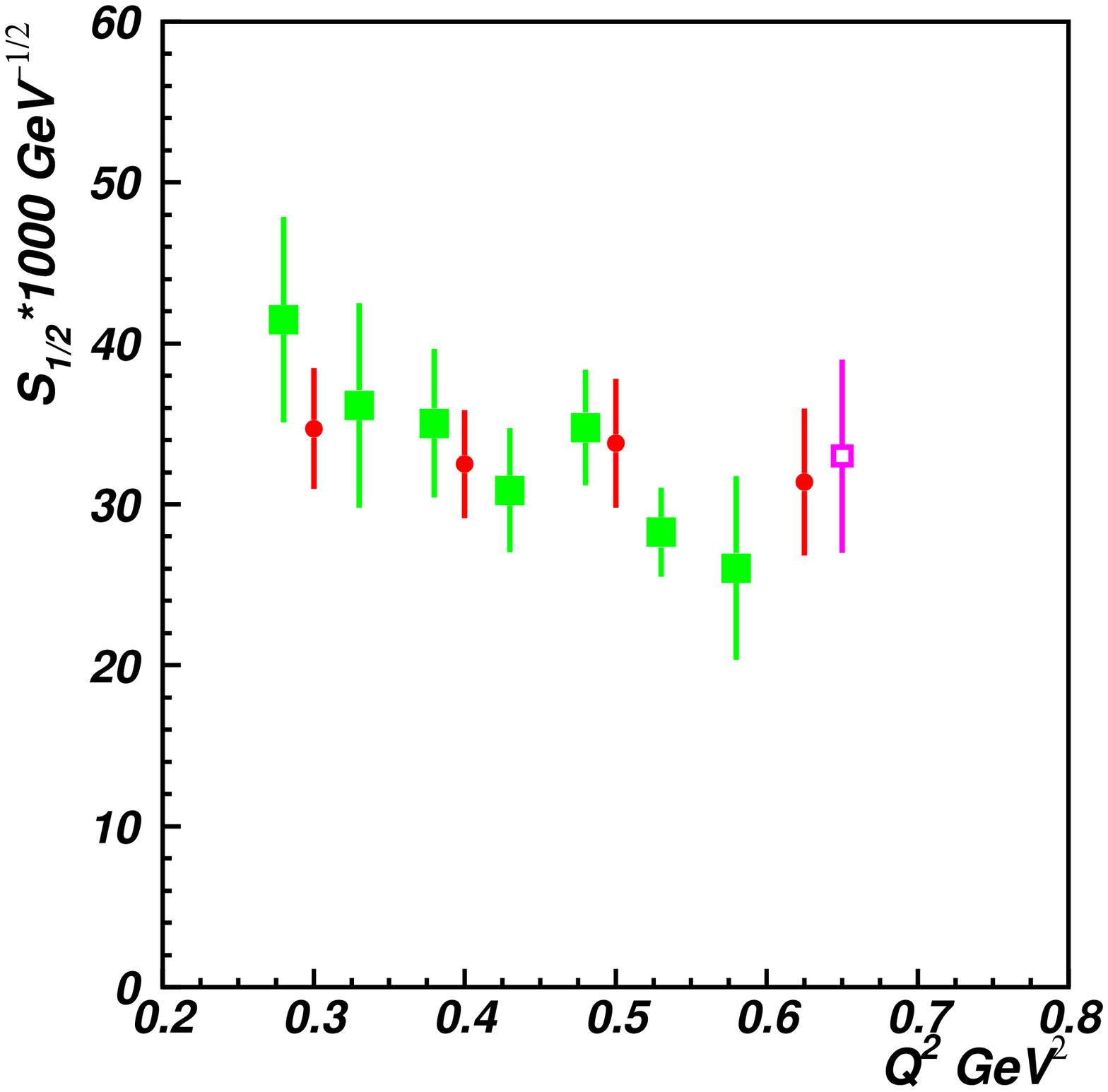}
\caption{\small $A_{1/2}$ (left) and $S_{1/2}$ (right) electrocouplings of 
$P_{11}(1440)$ state  determined from the CLAS data \cite{Fe09} on $\pi^+\pi^-p$
electroproduction (filled squares) and from $N\pi$ electroproduction 
\cite{Az09} (filled circles). Results from a combined analysis of 
$N\pi$ and $\pi^+\pi^-p$ channels \cite{Az05} are shown by open squares. Filled circle and triangle at the photon
point correspond to CLAS \cite{Dug09} and PDG  analyses of $N\pi$ channels, respectively.}  
\label{p11}
\end{figure}

\begin{figure}[htp]
\includegraphics[width=7cm]{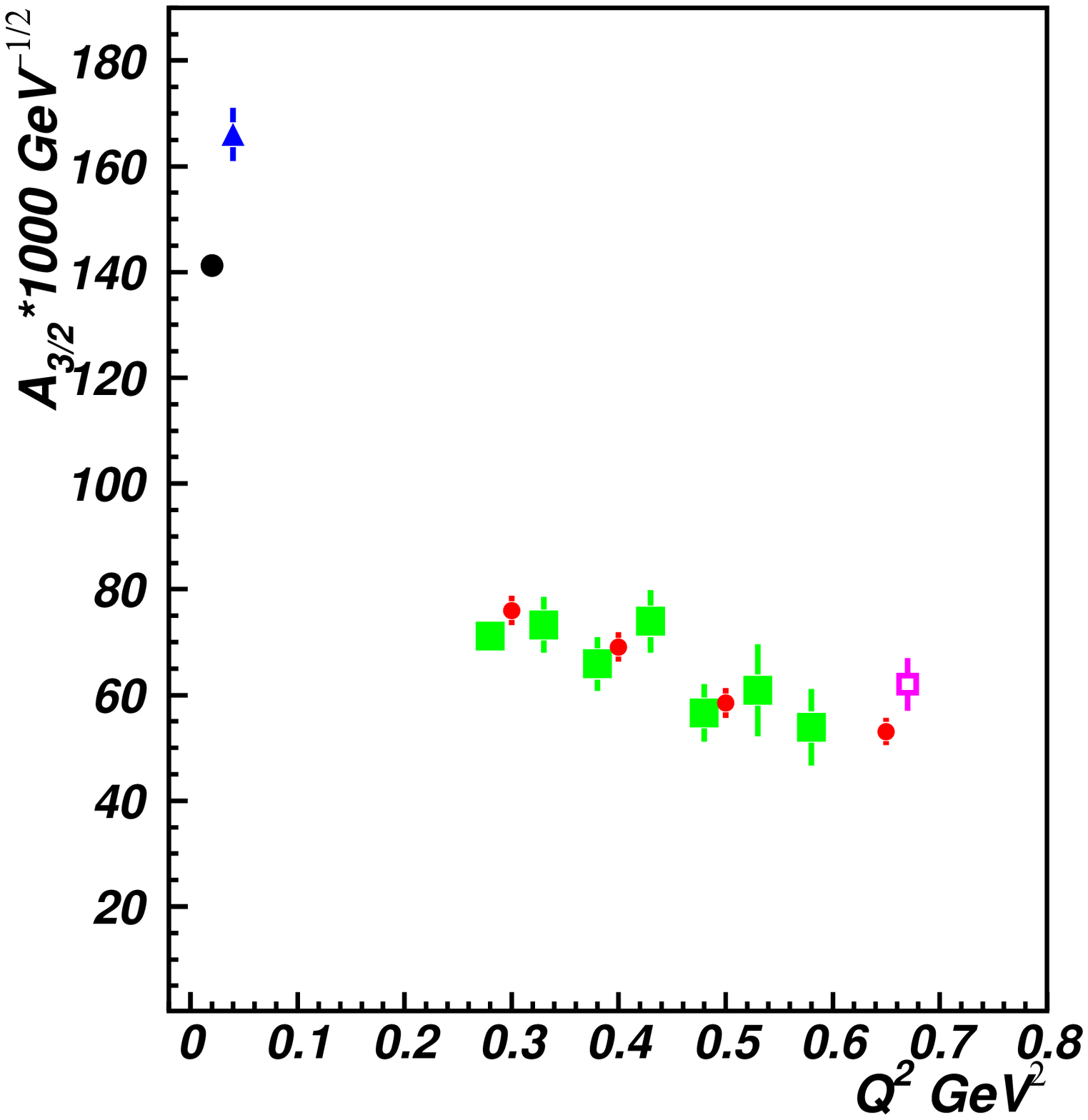}
\includegraphics[width=7cm]{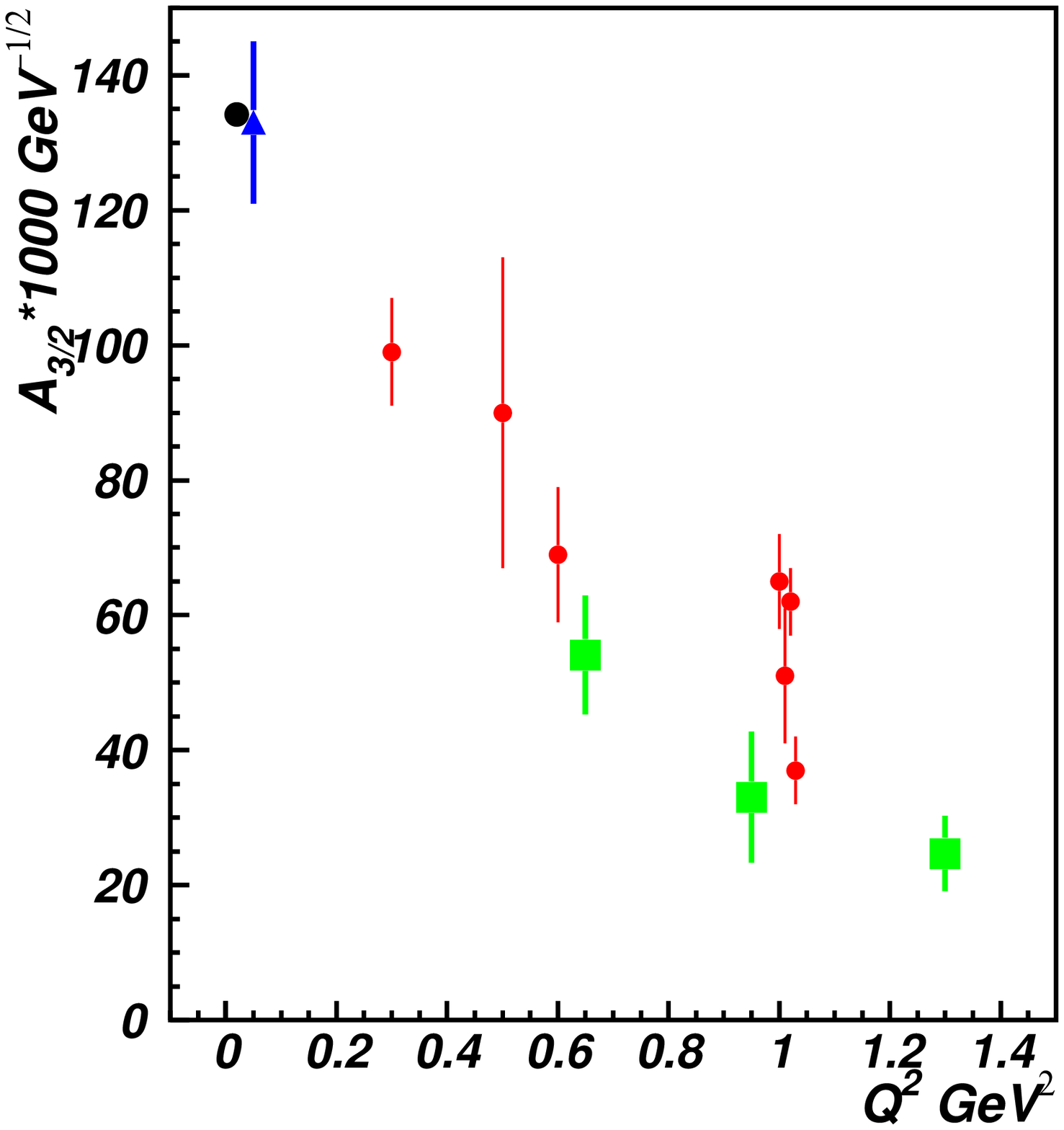}
\caption{\small The $A_{3/2}$ electrocouplings of 
$D_{13}(1520)$ (left) and $F_{15}(1685)$ (right) resonances. The results from analysis
of $\pi^+\pi^-p$ CLAS data \cite{Ri03,Fe09} are shown by filled squares.
Electrocouplings of $D_{13}(1520)$ state determined from analysis of the CLAS data
on $N\pi$ electroproduction \cite{Az09} and electrocouplings of $F_{15}(1685)$ state 
obtained from analyses of 
previously available world data, taken from  the compilation \cite{Bu03}, are shown by 
filled circles. Filled circles and triangles at the photon
point correspond to CLAS \cite{Dug09} and PDG analyses of $N\pi$ channels, respectively.}  
\label{d13f15}
\end{figure}

\begin{figure}[htp]
\includegraphics[width=6cm]{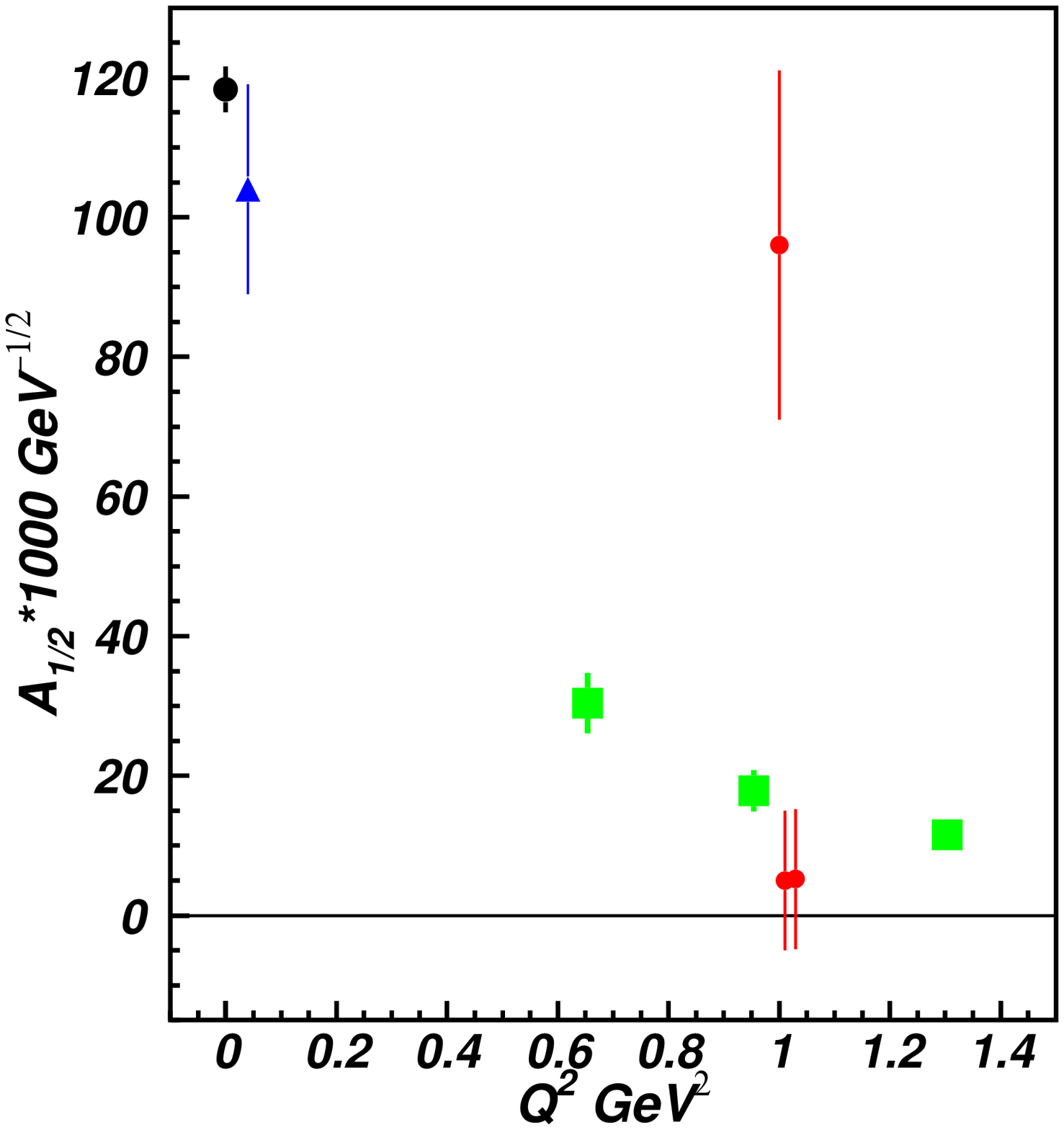}
\includegraphics[width=6cm]{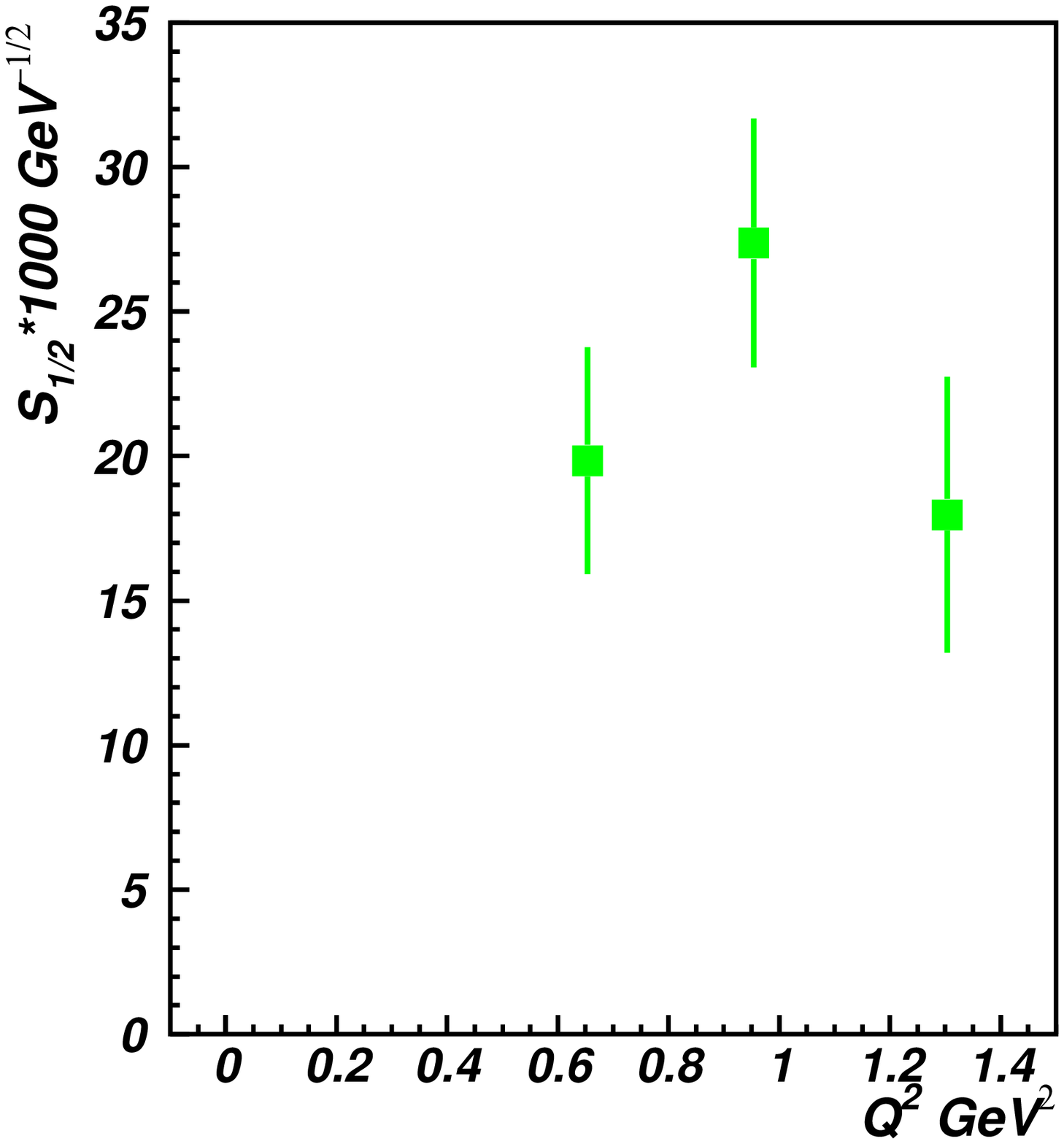}
\includegraphics[width=6cm]{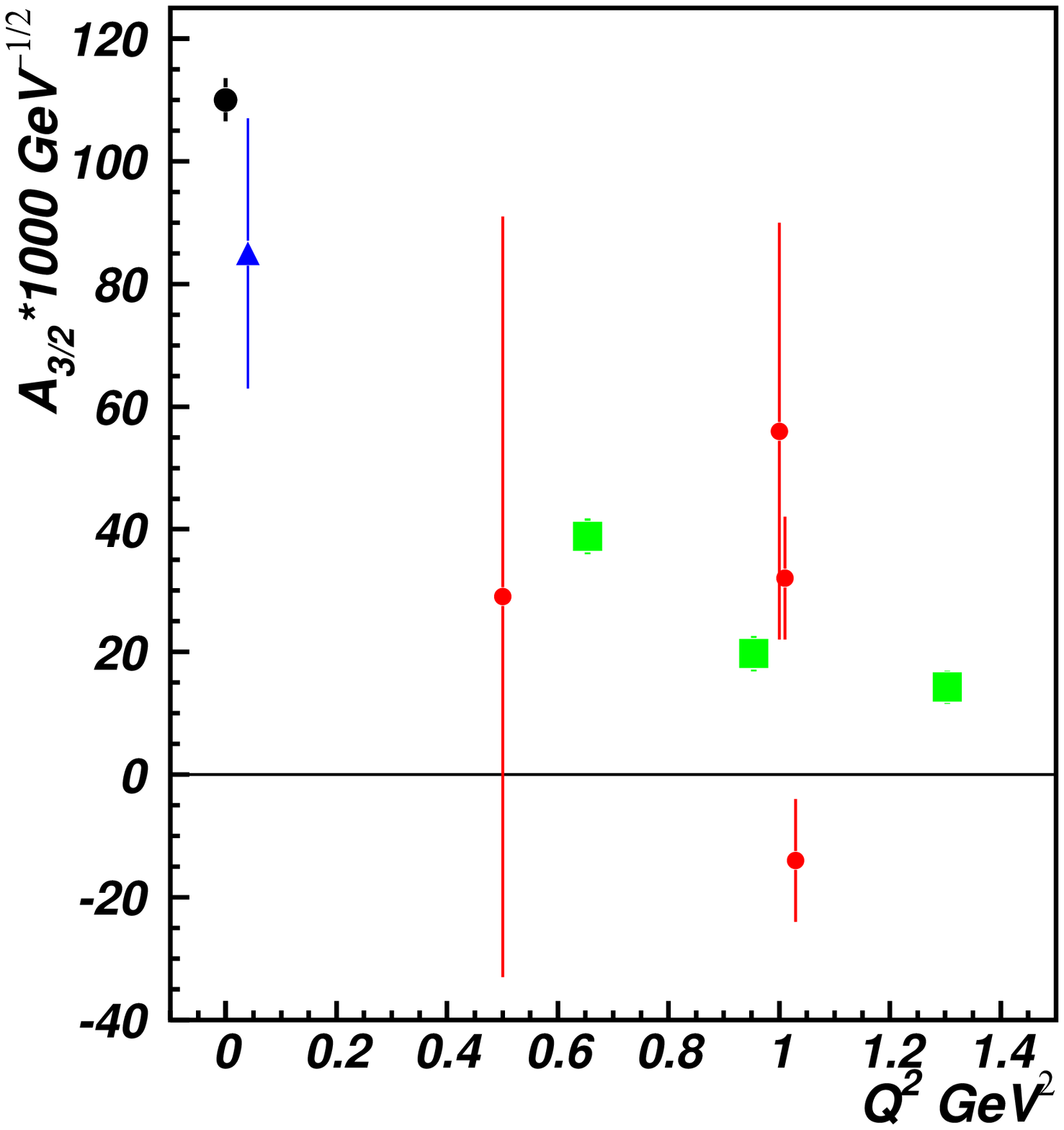}
\caption{\small Electrocouplings of 
$D_{33}(1700)$ resonance. Results from analysis
of $\pi^+\pi^-p$ CLAS data \cite{Ri03} are shown by filled squares, while filled circles 
represent the results obtained worldwide from analyses of $N\pi$ 
electroproduction data, taken from the
compilation \cite{Bu03}. 
 Filled circles and triangles at the photon
point correspond to CLAS \cite{Dug09} and PDG analyses of $N\pi$ channels, respectively.}  
\label{d33}
\end{figure}
 $\gamma_{virt}NN^*$ electrocouplings, as well as resonance $\pi \Delta$ and $\rho p$ partial hadronic
 decay widths assigned to selected in the fit calculated cross sections were averaged,
 giving us  the mean values of resonance parameters determined from the data fit.
 Dispersions in the sets determined in the fit resonance parameters were 
 considered as 
 their uncertainties.

\section{Results}
Electrocouplings of the $P_{11}(1440)$ state determined from the analysis are 
shown in Fig.~\ref{p11}.
The error bars include
quadratic sums of statistical and systematic uncertainties. We also show electrocouplings of 
this state obtained 
from a combined analysis of the data on $\pi^+n$ and $\pi^0p$
electroproduction channels \cite{Az09}. Electrocoupling values extracted 
from analyses of  these
exclusive channels are in good agreement. We also found consistent values for electrocouplings
of $D_{13}(1520)$ and  $F_{15}(1685)$ states
determined independently from the two channels.
 As an example, the $A_{3/2}$ electrocouplings of these states are 
shown in  Fig.~\ref{d13f15}.

$N \pi$ and $\pi^+\pi^-p$ exclusive 
electroproduction channels represent two major contributors to meson 
electroproduction in the
$N^*$ excitation region with completely different non-resonant mechanisms. 
The successful description of a large body of 
observables in these two exclusive channels using nearly the same 
electrocouplings of $P_{11}(1440)$, $D_{13}(1520)$  and $F_{15}(1685)$ states 
provides strong
evidence for a reliable evaluation of these fundamental quantities.

Many excited states with masses above 1.6 GeV preferably couple to 
$N\pi\pi$ final states. Therefore, analysis of  $\pi^+\pi^-p$ 
electroproduction  is of
particular importance to obtain reliable information on high lying $N^*$ 
states.

\begin{table}
\begin{tabular}{|c|c|c|}
\hline
~~~~~Breit-Wigner ansatz:~~~~~ & $~~~~~\Gamma_{\pi\Delta}$, MeV ~~~~~ & $~~~~~\Gamma_{\rho p}$, MeV ~~~~~\\
\hline
Regular & 1.5 $\pm$ 1.1 & 114 $\pm$ 12. \\
Unitarized & 10.9 $\pm$ 1.4 & 83 $\pm$ 3.3  \\
\hline
\end{tabular}
\caption{\label{hadrdec} $\pi \Delta$ and $\rho p$ partial hadronic decay widths of 
$P_{13}(1720)$ state determined from
analysis of the CLAS $\pi^+\pi^- p$ electroproduction data \cite{Ri03}, employing the regular \cite{Ri00}
and the unitarized \cite{Ait72} Breit-Wigner ansatzs for resonant amplitudes.}
\end{table}

Figure~\ref{d33} shows $D_{33}(1700)$ electrocouplings
determined in our analysis in comparison with previously
available world data from analyses of $N\pi$ electroproduction. Large 
uncertainties of previous world
data on $D_{33}(1700)$ electrocouplings are related to the fact that $N\pi$ 
is a minor decay channel (10-20\%) of this state, that couples dominantly to 
$N\pi\pi$ final states (80-90\%). 
Analysis of the  $\pi^+\pi^-p$  data \cite{Ri03} within the framework of JM model 
 allowed us to obtain  accurate information on electrocouplings of 
this and also of the $S_{31}(1620)$, $S_{11}(1650)$, $F_{15}(1685)$, and  
 $P_{13}(1720)$ states at photon virtualities 
 from 0.5 to 1.5 GeV$^2$.  
Moreover, longitudinal electrocouplings of these
resonances have become available from this analysis for the first time.

In order to examine the effects of resonant amplitude unitarization, we
compared resonance parameters determined within the frameworks of 
the regular \cite{Ri00,Mo01} and, described above,
unitarized Breit-Wigner ansatzs. We found that $\gamma_{virt}NN^*$ 
electrocouplings determined by employing 
the regular and the unitarized  formulations are consistent with each other. 
However, 
unitarization of resonant amplitudes affects
substantially $\pi \Delta$ and $\rho p$ partial hadronic decay widths of excited
proton states. 
An example
is shown in the Table~\ref{hadrdec}, where we compare the $\pi \Delta$ and 
$\rho p$ partial hadronic decay widths of $P_{13}(1720)$ state, obtained from 
the  $\pi^+\pi^-p$ data fit by employing the JM model, that incorporate the 
regular and unitarized Briet-Wigner ansatzs, respectively.

\section{Conclusion and outlook}

Accurate information on the $Q^2$-evolution of $\gamma_{virt}NN^*$ 
electrocouplings for many excited
proton states with masses less than 1.8 GeV and at photon virtualities up to 
1.5 GeV$^2$ have become available from CLAS data on 
$\pi^+\pi^-p$ electroproduction. These results open up new opportunities for
theory to explore confinement mechanisms in the baryon sector through their manifestation 
in the structure of excited proton states of various quantum numbers, as it was outlined in
\cite{Wp09}.  The analysis reported here covers the range of photon virtualities,  
where both meson-baryon and quark degrees of
freedom can be relevant. 
Our results on high lying $N^*$ electrocouplings for the first time make 
it possible 
to explore the transition from meson-baryon to quark
degrees of freedom  in the structure of excited proton states with masses above
1.6 GeV within 
the framework of dynamical coupled channel approaches 
under development in EBAC at Jefferson Lab \cite{Lee10}.
 


\end{document}